\documentstyle[11pt,aasms4]{article}

\newcommand{\msol} {\mbox{M$_{\odot}$}}

\newcommand{\Hz} {~\rm{Hz}}

\begin{document}

\slugcomment{submitted to Ap.J (August 15, 2000)}

\lefthead{Jernigan}
\righthead{kHz QPO and GW Emission}

\title{KiloHertz QPO and Gravitational Wave Emission
as the Signature of the Rotation and Precession of a LMXB Neutron Star Near Breakup}

\author{{J. Garrett Jernigan}$^{1,2,3}$}

\altaffiltext{1} {Space Sciences Laboratory, University of California, Berkeley, CA 94720-7450\\
e-mail: jgj@ssl.berkeley.edu}

\altaffiltext{2} {Eureka Scientific Inc., 2452 Delmar St., Oakland, CA 94602-3017}

\altaffiltext{3} {Little H-Bar Ranch, 903 Mustang Ct., Petaluma, CA 94954-8546}

\begin{abstract}

The basic theory of torque free precession (TFP) of the outer crust of
a neutron star (NS) as the signature of the approach to NS breakup
is a viable explanation of the uniform properties of kHz Quasi-periodic Oscillations (QPO)
observed in X-rays emitted by Low Mass X-ray Binary (LMXB) sources.
The theory outlined in this paper relates
the intrinsic properties of NS structure to the observed kHz frequencies.
The range of kHz frequencies and the observed quality factors (Qs) are also
explained by this simple dynamical model.
A scenario that begins with the melting of the inner crust of an LMXB NS
creates the conditions necessary for the generation of kHz QPO.
We suggest that a mechanism analogous to that proposed to explain giant
glitches in radio pulsars drives the dynamics of variations of the kHz QPO.
Furthermore, the theory provides a simple
explanation for the high Q of the unique millisecond X-ray pulsar SAX~J1808.4-3658, 
and also explains why it does not exhibit kHz QPO.
The theory relates the ratio of the observed kHz frequencies to the
ratios of the components of the moments of inertia of the NS, 
thereby tightly constraining the equation of state (EOS) of NS matter (polytrope index $n\approx1.0$).
The TFP model is in strong contrast to existing models which primarily
relate the kHz QPO phenomenon to the physics of gas dynamics near the inner edge
of the accretion disk and the transition flow onto the surface of the NS.
The TFP theory is consistent with the presence of Lense-Thirring precession
of matter orbiting the NS. We suggest the possibility of the direct detection
of very low frequency ($\sim 1$ kHz) radio waves from magnetic dipole radiation and
also predict kHz gravitational wave emission from the LMXB Sco~X-1
that may be detectable by LIGO. 
The high accretion rates consistent with the predicted GW emission indicate the
likely conversion of some LMXBs to maximally rotating Kerr black holes (BH)
and further suggest that these systems are progenitors of some gamma-ray bursts (GRB).

\end{abstract}


\section{Introduction}

KiloHertz (kHz) oscillations in low-mass X-ray binaries (LMXB) were
first discovered (\cite{klis96}; \cite{strohmayer96})
in observations with the Proportional Counter Array (PCA)
onboard the {\it Rossi X-ray Timing Explorer} (\cite{bradt93}; \cite{swank97}).
The kHz Quasi-Periodic Oscillations (QPO) have now been observed in
the persistent flux of well over a dozen sources
(\cite{klis00}; and references therein).
These kHz QPO sometimes occur in pairs whose
difference frequency is not constant (\cite{klis97}; \cite{mendez98}; \cite{ford98})
and show properties that are remarkably uniform over a wide range of
accretion rates (\cite{psaltis98}).
Some of these sources also show oscillations in the $\sim300-600\Hz$ range that occur
during Type~I (thermonuclear) bursts (\cite{strohmayer98}).
Current models for kHz QPOs have explanations related to the inner
portions of the accretion disk and the transition flow of matter onto the
NS surface (\cite{miller98}; \cite{cui00}).
Other alternatives focus on the motions of blobs of matter near
the last stable orbit of a NS (\cite{zhang97})
and propose evidence for Lense-Thirring precession of orbiting matter (\cite{stella98}).
The unique millisecond X-ray pulsar, SAX~J1808.4-3658, shows
high Q millisecond pulsations ($401~{\rm Hz}$) and a lack of kHz QPOs (\cite{wijnands98b}).
Also, pairs of kHz QPO are not observed in sources with black holes (\cite{klis00})
indicating that the generation of kHz QPO is likely related to a mechanism 
near the surface of the NS.

In this paper, we present the torque free precession (TFP) model
for kHz QPO observed in LMXB sources which
relates the properties of the observed X-ray emissions from the surface
of the NS over a wide range of accretion rates
to the intrinsic properties of neutron star (NS) structure.
This model is in strong contrast to existing models which primarily
relate the kHz QPO phenomenon to the physics of gas dynamics
external to the NS surface.

\section{Torque Free Precession Model}

KiloHertz oscillations are explained as a natural consequence of the spinup
of a neutron star to near breakup due to sustained high rates of accretion.
In this model, the fluid core of the NS spins up due to accretion torques ($\sim300-800\Hz$).
Further transfer of angular momentum to the NS causes the ``crust" of the NS
to partially decouple from the core and rotate at a higher rate (up to $\sim1100\Hz$)
than the fluid core. The crust of the NS both rotates and undergoes torque free
precession (TFP) as a rigid body: this causes a modulation of the X-ray emission
from this observable outer region at a pair of frequencies in the kHz band.
We designate these frequencies as $\nu_{\ell}$ and $\nu_u$, the lower and upper kHz QPO.
These oscillations are quasi-periodic because of fluctuations in the
moments of inertia of the outer region of the NS and possible exchange of angular
momentum between the rigid crust and the fluid core.
Although the kHz QPOs observed in X-rays are the most well studied predictions of this
model, most of the energy released is likely in the form of kHz gravitational waves (GW).

\subsection{Theoretical Overview}

The crust of the NS will undergo classical TFP in the absence of any accretion.
Torques due to the accretion flow or changes in the internal structure of the NS
will modify the precession and rotation of the NS on a timescale which is slow
compared to either the spin of the crust at frequency $\nu_{cs}$,
the precession of the crust at frequency $\nu_{cp}$, or the spin of the fluid core at a
frequency $\nu_{fs}$.
For the purposes of the model proposed here, we identify the ``rigid" crust as that portion
of the outer part of the NS which undergoes rotation and precession independently
of the fluid core of the NS.
The structure of the outer layers of a NS is
divided into a surface layer at low density ($<10^{6} {\rm~gm~cm^{-3}}$),
an outer crust ($\sim10^{11} {\rm~gm~cm^{-3}}$),
and an inner crust ($\sim10^{14} {\rm~gm~cm^{-3}}$).
These layers are outside of an neutron fluid core at nuclear density ($\sim10^{15} {\rm~gm~cm^{-3}}$).
There may be an interior core region that is solid or some other form of matter
distinct from a neutron liquid at even higher densities (\cite{shapiro83}).
The exact details of the NS structure are not important for the simple application of the TFP model.

Simplifying these constructs,
we refer to only two components, the ``crust" and the ``core" of the NS.
In a latter section, we identify the outer crust as the portion of the NS that
undergoes TFP because the inner crust is likely melted in an accreting LMXB NS. 
We further assume that the crust is rigid and undergoes rotation and precession
while the core is a fluid and only undergoes rotation. Clearly there is a complex interface
between the crust and core as well as time dependent deformations of the crust and core,
and perhaps non-rotational differential flows in the fluid interior. All of these details
are ignored in modeling the basic response of the NS to increasing angular momentum
and approach to breakup conditions.

Let $I_{c\parallel}$ and $I_{c\perp}$ be the parallel and perpendicular components
of the moment of inertia of the crust of the NS as related to the rotational axis of the NS.
We approximate the response of the crust of the NS to a variation in rotation rate
by the following pair of equations where $\nu_{cs}$ is the NS crust spin frequency
and $I_{c0}$ is the moment of inertia of the non-rotating crust. Clearly,
$I_{c\parallel}$ must equal $I_{c\perp}$ in the limit of no rotation:

\begin{equation}
I_{c\parallel} = I_{c0} \left[ 1~+~ {\nu_{cs} \over \nu_{c\parallel}} \right]
\label{Iparallel}
\end{equation}

\begin{equation}
I_{c\perp} = I_{c0} \left[ 1~+~ {\nu_{cs} \over \nu_{c\perp}} \right]
\label{Iperp}
\end{equation}

where $\nu_{c\parallel}$ and $\nu_{c\perp}$ are constants that characterize
the response of the crust to rotation,
in the directions parallel and perpendicular to the NS spin axis.

We also define a dimensionless ratio ellipticity $\epsilon_c$: 
\begin{equation}
\epsilon_c = {{\left[ I_{c\parallel} - I_{c\perp} \right]} \over {I_{c\perp}}}
\label{epsilon_c}
\end{equation}

The ellipticity $\epsilon_c$ is related to the eccentricity $e_c$ by

\begin{equation}
\epsilon_c = {\nu_{cp} \over \nu_{cs}} = {e_c^2 \over {2~-~e_c^2}} = {\nu_{\ell} \over \nu_u}
\label{epsilon}
\end{equation}

where the eccentricity $e_c$ is defined as
\begin{equation}
e_c^2~=~1~- {\left( r_p \over r_e \right)}^2
\label{radius}
\end{equation}
Here $r_e$ indicates the equatorial radius of the NS and $r_p$ indicates the
polar radius. The NS is assumed to have the shape of a oblate spheroid which
is induced by rotation of the crust at frequency $\nu_{cs}$.
Note that in equation \ref{epsilon}
we have made the interpretation that the observed lower kHz QPO frequency is
the crust's precession frequency ($\nu_{cp}=\nu_{\ell}$) while 
the upper kHz QPO frequency is the crust's spin frequency ($\nu_{cs}=\nu_u$).
Equations \ref{epsilon_c} and \ref{epsilon} are consistent with TFP of a rigid body (\cite{goldstein50}).

From the Maclaurin sequence of the equilibrium figures of uniform rotation and density (\cite{chandra69})
critical values of the ellipticity $\epsilon$ are determined as purely
geometrical functions of the ratios of moments of inertia. 
In the following discussion $\epsilon$ and $e$ (no subscripts) refer to the ellipticity and
eccentricity respectively of a MacLaurin spheroid defined analogously as the equations
\ref{epsilon_c}, \ref{epsilon} and \ref{radius} for the crust.
Generalizations (\cite{ostriker73}) of the Maclaurin sequence indicate that the critical
values of $\epsilon_c$ are approximately the same as those for $\epsilon$,
even in cases of non-uniform rotation and density.
Even if the surface shape of the NS does not match an ideal oblate spheroid, $\epsilon_c$ is
still well defined as a function of the ratios of the moments of inertia. The critical 
values of ellipticity ($\epsilon$) and also the eccentricity ($e$)
are often expressed as corresponding critical values
of the ratio of kinetic energy ($T$) to the absolute
value of the potential energy ($\left| W \right|$)
designated $T \over {\left| W \right|}$.

Considering a MacLaurin sequence of increasing angular momentum, we first encounter a secular instability
at $\epsilon=0.4930;~e=0.8127; { { T \over {\left| W \right|} } = 0.1375 }$.
This first instability determines the maximum rotation of a NS for an assumed equation of state (EOS) and mass.
Note that the values of eccentricity (and also $T \over {\left| W \right|}$) are slightly less than
the critical value for the ``normal" sequence of maximally rotating NS computed by Cook
(see table 7 of \cite{cook94b}). This result is robust for all EOS for which Cook
computed models.
The next important critical value 
($\epsilon=0.7618;~e=0.9300; { { T \over {\left| W \right|} } = 0.2379 }$)
occurs at the state with the maximum possible rotation
rate. Beyond this point, increasing the angular momentum will not increase the rotation rate.
This state does not indicate the onset of any instability, but rather
reflects the fact that the system has reached its maximum possible spin for a rigid internal structure.
The next critical value
($\epsilon=0.8315;~e=0.9529; { { T \over {\left| W \right|} } = 0.2738 }$)
occurs when the rotating rigid body becomes dynamically unstable because of the rapid growth  of
nonradial toroidal modes. Any rotating system, including a rigid body,
is unlikely to maintain a value of eccentricity greater than 0.9529.
 
Recall that we have previously associated the 
precession frequency of the NS crust ($\nu_{cp}$) with the
observed lower kHz QPO frequency ($\nu_{\ell}$), and the 
spin frequency of the crust 
($\nu_{cs}$) with the upper kHz QPO frequency ($\nu_u$).
For the conditions considered here the accretion is likely higher onto a large equatorial
region of the crust. The polar regions of the crust likely rotate with
interior fluid which is observed in Type~I bursts from LMXB NSs
(see a later section for more discussion of this possibility).
At low angular momentum, the fluid and crust rotate together as a single rigid body
at the spin frequency $\nu_{fs}~(=\nu_{cs})$.
As the angular momentum is increased, the NS reaches the critical internal state which
results in the partial decoupling of an equatorial portion of the crust from the interior of the NS.
The details of this internal change in the NS structure are likely due to heating
of the inner crust due to nuclear burning associated with the accretion
of matter onto the NS surface and subsequent migration to deeper layers of the NS.
This catastrophic event will occur at a unique spin for a particular NS among an uncertain range of
spin frequencies depending on the details of the history of accretion.
Since we have no quantitative theoretical model we identify a plausible
range as $300-800\Hz$ for consistency with observations of NCO in Type~I bursts
and the possibility that Z-sources have spins as high as $800\Hz$.
The details of this event are discussed in a later section. For the discussion here,
we only need assume that a portion of the crust has been reduced in mass ($\sim10^{-4}\msol$) and
is less coupled to the NS fluid interior. We will show later that the EOS of the NS is
such that at the time just prior to the catastrophic event, the shape of the NS has an eccentricity
$e\sim0.2$ and is a stable. Subsequent to the decoupling of the outer crust from the interior
of the NS, this part of the crust begins to rotate at a spin frequency different and higher than
the fluid core. In this context, the descriptive term decouple means that the interaction of
the crust (reduced in mass) with the NS interior is sufficiently weak that an equatorial portion of the
crust can rotate and precess independently from the NS fluid core. Clearly there is
still a significant interaction between the crust and core.
The crust will experience torques due to both accretion and internal changes in the NS.
Since the crust is reduced in mass, its spin can more easily be changed. The torques
due to accretion are not large enough to change the spin of the crust by a large amount
($\sim300$ to $\sim1200\Hz$). However, internal changes in the NS will produce torques on the
crust that can significantly spin the crust up or down. The crust will always spin faster than
the fluid core. Therefore the spin frequency of the fluid core ($\nu_{fs}$) is less than the
minimum value of the frequency of the upper kHz QPO ($\nu_u$) in each particular source.
As the crust spins up its shape will change due to a continuous elastic
response. As the crust reaches a spin frequency of $\sim500\Hz$ and an eccentricity
$e\sim0.8127$ it approaches the onset of a secular instability.
The response of the crust to this secular instability is the likely initiation of TFP.
This transition near $\sim500\Hz$ is supported by observations of 4U1728-34 and 4U1608-52
that show both kHz QPO for $\nu_u>\sim500\Hz$
but only the upper kHz QPO for $\nu_u<\sim500\Hz$ (see figure 1 of \cite{mendez00}).
The crust now both spins and precesses over a range of spin frequencies
$\nu_{cs}\approx 500-1100\Hz$ ($e_c\approx0.8-0.93$), 
while the fluid core continues to rotate near the same rate that it had
just prior to decoupling of the equatorial crust.

The NS does not ``break up" as a single body but rather breaks into two separate bodies.
The core is a fluid which rotates but does not precess, while the ``rigid" crust  
rotates and precesses. This motion of the crust leads to the appearance of the pair of kHz QPO
at the rotation frequency of the crust $\nu_{cs}~(=\nu_u)$ and the precession frequency of the
crust $\nu_{cp}~(=\nu_{\ell})$. As $\nu_{cs}$ increases to 
$\sim1100\Hz$, the crust reaches the next critical value of eccentricity
$e\sim0.9300$ which corresponds to the maximum rotation rate of the rigid crust that
will allow both rotation and precession to occur. As the rotation rate increases higher than $\sim1100\Hz$
toward slightly greater values ($\sim1200\Hz$ which corresponds to $e\sim0.9529$),
the crust approaches dynamical instability which quenches its precession.
An alternative and perhaps more likely scenario is that for spins above $\sim1100\Hz$ the
equatorial crustal material and the plasma atmosphere above the crust are moving very close to escape
velocity of the NS and may merge with orbiting material at the inner edge of the accretion disk.
The upper kHz QPO in the frequency range $\sim1100-1200\Hz$ may be caused by orbital motion
hence the lack of TFP and a missing lower kHz QPO. When this spin/orbital frequency reaches $\sim1200\Hz$
the accretion disk and crust come into contact which results in the cessation of the upper kHz QPO.

At this state of maximum angular momentum, the fluid core
is still rotating at frequency $\nu_{fs}$ (some value in the range $\sim300-800\Hz$)
while the crust is rotating at $\nu_{cs}$ ($\sim1100\Hz$).
For greater angular momenta, the entire NS system (crust and core) can no longer rotate as two
bodies. Also the surface of the crust is moving at a speed near the escape velocity. Further increases
in angular momentum are likely to lead to mass ejection or the
real ``breakup" of the NS. The decoupling of the crust from the core at spin frequencies
$\sim300-800\Hz$ stalls the further spinup of the NS while the equatorial crust spins up to $\sim1100\Hz$.
At this point the NS (crust and core) has reached the breakup state. 

\subsection{TFP Model Calculation}

The successful quantification of the TFP model is confirmed by fitting the observed frequencies
of the kHz QPO from Sco X-1. We can approximately model the moments of inertia of the crust
($I_{c\parallel}$ and $I_{c\perp}$) as a response
to rotation at frequency $\nu_{cs}$ as a linear increase from an initial value at zero rotation
($I_{c0}$) as shown in equations \ref{Iparallel} and \ref{Iperp}.
The fixed free parameters in this model are $\nu_{c\parallel}$, $\nu_{c\perp}$ and
$I_{c0}$. The value of $\epsilon_c$ defined by equation \ref{epsilon_c} is independent
of the value of $I_{c0}$ since it only depends on ratios of moments of inertia.
Equation \ref{epsilon} relates the moments of inertia to the ratio of the
lower and upper frequencies of the kHz QPO with the interpretation that the
upper frequency is the spin frequency of the crust ($\nu_u=\nu_{cs}$) and the lower frequency
is the precession frequency of the crust ($\nu_{\ell}=\nu_{cp}$). 
We use this model to fit the data for ScoX-1 and thereby empirically determine the
values for $\nu_{c\parallel}=1009.4\pm1.0\Hz$  and $\nu_{c\perp}=6435\pm22\Hz$.
These values are computed by a $\chi^2$ fit to the 38 pairs of measurements of $\nu_l$ and $\nu_u$
for Sco~X-1 with two free parameters and $50\%$ error bars ($\Delta \chi^2=2.37$)
derived according to \cite{lampton76}.
The minimum value of $\chi^2$ is 92.4 for 34 degrees of freedom. 
This fit is of similar quality to that from \cite{psaltis98} and implies systematic errors
$\sim2$ times larger than the statistical errors.
The value of $I_{c0}$ is not determined by the fit.

This fit is shown in figure 1 (a) and (b). Panel (a) shows the fit to the data
by the TFP model (solid line) as well as 
the phenomenological equation of \cite{psaltis98} (dashed line). 
The fits are similar; however for the TFP model we are fitting a
dynamical model rather than a phenomenological function.
In panel (a), the data point near the abscissa value of $500\Hz$
is obtained from observations of GX~5-1 (\cite{wijnands98b}). It has
not been included in the fit for $\nu_{c\parallel}$ and
$\nu_{c\perp}$,
but is included in the figure to demonstrate that   
the TFP model appears to fit the data over the
full observed range of $\nu_u$ ($500-1100\Hz$) with $\nu_l$ present.
In panel (b) we show an expanded view of the Sco~X-1 data and the 
models.
Panels (a) and (b) show the ordinate as the difference frequency
of the kHz QPO as usually shown in prior work (\cite{psaltis98} and references therein).

In panels (c) and (d) we show the same data and the 
TFP model fit with the ordinate as
the ratio $\nu_{\ell}/\nu_u$ which is equal to the ellipticity of the crust $\epsilon_c$. 
Again the TFP model is the solid curve which is shown to match
the Sco~X-1 data and the single point from GX5-1. Panel (d) is the same as panel (c)
at an expanded scale showing a better view of the Sco~X-1 data.
Panel (c) also shows horizontal dotted lines for the critical values of the
eccentricity of the crust. Also shown in panel (c) is one curve that correspond to
a particular model of a rotating NS.
The model uses the specific EOS FPS (\cite{lorenz93}) for a NS of mass $1.4~\msol$ taken from
a computation in \cite{cook94b}. Given the interpretation of the model proposed here,
the curve for the real EOS lies below the solid curve for all values of $\nu_u$.
By inspection of table 15 from \cite{cook94b}, it can be shown that the ratio of
the radius of a NS at maximum spin is close to $\sqrt2$ times the radius at zero spin.
Similarly the decoupled crust also increases its equatorial radius by $\sim \sqrt2$
as it approachs maximum spin. This result is consistent with the condition $\nu_u~\sim~\nu_{c\parallel}$
which from equation \ref{Iparallel} means that the moment of inertia of the crust has
doubled corresponding to an increase in the equatorial radius by a $\sim \sqrt2$.
In other words, the radius of the decoupled crust at maximum spin is nearly the
same as the radius that the NS would have at maximum spin if the crust did not decouple.
The condition that the rotating and precessing crust must be outside the radius of the
fluid core  mean that the EOS of the NS must have a maximum spin greater than $\sim1100\Hz$.
Since observations support the contact on the inner edge of the accretion disk with
the equatorial crust at maximum spin of the crust $1100-1200\Hz$ therefore the EOS of the
NS must have a maximum spin less than $\sim1200\Hz$ and likely close to $\sim1100\Hz$.
Of all the EOS considered by \cite{cook94b}, EOS FPS is closest
to matching this requirement. Of the polytropes considered in \cite{cook94a} only
the index $n\sim1.0$ matches this requirement.
If the TFP model is correct, then the EOS of a NS is tightly constrained
within a range that is comparable to the systematic errors in the computation
of the structure of a rotating NS.
The best fit polytrope for the interior of a NS
has an index $n$ likely within $\pm0.1$ of $1.0$.
The empirical relationship in equation \ref{polytrope} relates the value of the
polytropic index $n$ to the maximum spin frequency $\nu_{max}$ of a NS with a mass ${\rm M_{ns}}$.  
This equation is derived from equation 31 and the data in table 3 from \cite{cook94a}.
For a $1.4\msol$ NS and assumimg $\nu_{max}$ is equal to the maximum value of $\nu_u\sim1100\Hz$
with a simultaneous measurement of $\nu_{\ell}$ yields a value $n\sim0.94$.
If the masses of the NS in an LMXB are as high as $\sim2~\msol$ then $n\sim1.16$.
The basic conclusion of a tightly constrained EOS with $n\sim1$ still holds.

\begin{equation}
n~=~1.0~+~1.4 \left[ log_{10} \left( {\rm M_{ns}} \over 1.4~\msol \right) - log_{10} \left( \nu_{max} \over 1004\Hz \right) \right]
\label{polytrope}
\end{equation}

In panel (d) we show an expanded view of the Sco~X-1 data.
Note that the maximum value of $\nu_u$ at $\sim1070\Hz$ corresponds to the critical
value of $e\sim0.9300$. The dashed curve in panels (c) and (d) shows the locus for a
Maclaurin spheroid that just passes through these points at maximum spin with precession.
The mass ($1.4~\msol$) and radius ($\sim17~$km) of the spheroid is approximately correct
for nominal values of a NS.
The upper kHz QPO  from Sco~X-1 has been observed at values greater than $1100\Hz$
but without the presence of the lower kHz QPO. The model suggests that the maximum
spin rate without precession present (missing lower kHz QPO) would occur at a frequency
$\sim1180\Hz$ which is determined by the intersection of the $e=0.9529$ line
with the solid curve. This point marks an upper limit for the breakup frequency of the NS in the
context of the TFP model. An alternative, and more likely interpretation, is that the
true breakup frequency of full NS (core and crust) is closer to $1100\Hz$. The observed
upper kHz QPO in the frequency range $\sim1100-1200\Hz$ (without the lower kHz QPO)
marks a transition region in which the outer crust and the NS plasma atmosphere are
moving near escape velocity and are merging with the inner edge of the accretion
which has a matching Keplerian velocity. Full merger of the outer crust
with the inner edge of the accretion disk marks the cessation of the upper kHz QPO.

In panel (c) a vertical dot-dash line is plotted at a frequency $\nu_u=401\Hz$,
which is the spin frequency of SAX~J1808.4-3658, the millisecond X-ray pulsar discovered 
by \cite{wijnands98a}. The locus for EOS FPS intersects
the line for this pulsar near $\epsilon_c\sim0.03$.
For the EOS FPS the radius of the NS varies from $10.85-15.45$ km
(see table 12 in \cite{cook94b}).
For a $\sim1.4~\msol$ NS with $\epsilon_c=0.03$ (or $e=0.17$) the radius as set by
EOS FPS is $\sim11~$km in agreement with the upper limit $\sim16~$km suggested by \cite{burderi98}.
A high Q pulsar such as SAX~J1808.4-3658 has not reached the condition for melting the
inner crust. In the context of the TFP model such a high Q pulsar would never show kHz QPO. 
Also the millisecond radio pulsar PSR~1937+214 discovered by \cite{backer82} has a
spin frequency of $\sim640\Hz$. The high Q of this pulsar is consistent with the lack of accretion
and a solid inner crust. Perhaps this radio pulsar once had a melted inner crust and
significant accretion rate and has since reformed a solid inner crust after cooling
subsequent to a cessation of accretion.

\section{Scenario for Initiation and Maintenance of TFP}

The correctness of the TFP model is strongly supported by the 
the consistent prediction of the sequence of kHz QPO observed in LMXB sources.
The angular momentum transfer via accretion is
too small to drive a large variation in the spin of the outer crust ($300-1200\Hz$)
over a short period ($\sim day$). 
Any realistic and complete TFP model requires a mechanism that taps 
a small fraction of the large reservoir of both
rotational energy and angular momentum of a rapidly spinning NS that drives the outer
crust to vary in spin frequency.
We suggest that the controlling mechanism is related to giant glitches
observed in radio pulsars (\cite{alpar81}).

Initially the NS is spinning up due to the accretion of material onto the surface
of the outer crust. Over a long accretion timescale the outer and inner crust are repeatedly
replenished. \cite{brown00} has calculated detailed models for the equilibrium structure of
the crust in such a accreting source.
%The melting temperature of the crust is
%given by equation \ref{melt} (\cite{ushomirsky00}).
%\begin{equation}
%T_m = I_{c0} \left[ 1~+~ {\nu_{cs} \over \nu_{c\parallel}} \right]
%\label{melt}
%\end{equation}
For a sufficiently high accretion rates ($10^{-9}\msol yr^{-1}$), Brown's
model predicts that the temperature in the inner crust is independent of the
surface temperature of the NS. Just below the layer beginning at onset of neutron drip,
the boundary between the outer and inner crust, the nuclear reactions heat the
interior of the inner crust. Most of this heat is conducted into the core of the NS.
Brown's model suggests that the inner crust will melt for a sufficiently high
accretion rate. The fall of the average value of Z in the inner crust also lowers
the melting temperature of this layer as well. Also the presence of any r-mode
driven by the rapid spin of the NS will heat the boundary-layer between the fluid
core and inner crust (\cite{bildsten00}) and further raise the temperature of the inner crust.
This scenario leads to the possibility of melting the inner crust. However for rates as high
as 5 times Eddington the outer crust may not melt (see figure 10 of \cite{brown00}).
The conductivity of the outer crust is low compared to that for the inner crust
(see figure 5 of \cite{brown00}).
We suggest that the onset of TFP occurs when the inner crust melts leaving the outer crust
as a solid. The spin of the NS star at the time when the inner crust melts will vary
depending on the history of the accretion rate, the detailed structure of the inner and outer
crust, and the presence of any additional non-nuclear heating sources (for example an r-mode).
The onset of melting of the inner crust is consistent with a range of spin
rates of the fluid core seen in Type~I bursts ($300-800\Hz$). In the TFP model only the
equatorial portion of the outer crust of the NS spins up and down ($300-1200\Hz$).
Large polar regions of the NS outer crust continue to spin with the interior fluid core.
A later section contains a more detailed discussion of the implications of the TFP model
for observations of Type~I bursts in LMXB sources. 

Once the inner crust melts then the outer crust can undergo TFP
driven by a mechanism that taps internal NS rotational energy.
We suggest that a
likely candidate for this mechanism is similar to that proposed to explain giant
glitches in radio pulsars (\cite{alpar81}). These giant glitches are observed to cause
discrete jumps in the spin period of the outer crust that correspond to a changes
in angular momentum of the a large portion of the NS. The size of the giant glitch in Vela
is $\sim 10^{-6}$ expressed as the fractional change in the spin frequency before and after the glitch.
The energy source for the giant glitches is likely a internal superfluid component which has 
a moment of inertia, $\sim10^{-2} I_{f\parallel}$,
as measured by the fractional change of the spin's first derivative before and after the giant glitch,
where $I_{f\parallel}$ is nearly the total moment of inertia of the NS.
Both of these parameters are determined by radio observations of
giant glitches in the spin of a NS and are not based on any theoretical
models of NS interiors (\cite{alpar81}).
The energetics of this "glitch" in
the spin rate of an LMXB NS is somewhat larger than a giant glitch in a radio pulsar.
This possibility is not that extreme since a LMXB NS is spinning $\sim50$ times faster than the
radio pulsars (Vela) for which the giant glitches are measured.
We suggest that the "glitch" in an LMXB likely scales as the period of rotation
yielding an estimate of the fractional change in the total angular momentum $\sim10^{-4}$.
If the superfluid transfers a large fraction of this stored angular momentum
to the solid outer crust ($I_{c\parallel}\sim 10^{-4}I_{f\parallel}$)
then a small differential in the spin of the superfluid ($\sim5\Hz$) compared to the
spin of the core of the NS ($300-800\Hz$) would cause a large change in the spin of the
outer crust (from $300-800\Hz$ to $\sim1100\Hz$) since the ratio of the moment of inertia of the
outer crust to that of the superfuild layer is $\sim 10^{-2}$.
Also the timescale for recovery from a giant glitch in radio pulsars is order $\sim100~{\rm days}$.
This long timescale is likely due to low viscosity in the superfuild layer.
Theoretical estimates for this timescale are uncertain but may also scale with the period of rotation
(see equation 36 in \cite{sauls88}). Therefore in an LMXB NS spinning $\sim50$ times faster than
the Vela pulsar, the glitch recovery time would become $\sim 1~{\rm day}$.
We suggest that the outer crust spins up and down rapidly ($300-1100\Hz$) on a timescale of
${\rm days}$ due to the pinning and unpinning of vortices in a superfluid layer within the NS.
This superfluid layer (which includes the melted inner crust) likely has an interface at the base
of the outer crust where it interacts via slow viscous effects. This allows the outer crust to
undergo TFP on a fast timescale ($1~{\rm ms}$) while exchanging angular momentum with the superfluid reservoir
on a slow timescale ($\sim$hours). This exchange of angular momentum proceeds in a limit-cycle
that is observed in the spin and precession of outer crust. This process results in a rapid changes in the
spin and shape of the outer crust and is consistent with the observed Qs of the kHz QPO.
The random meander of this limit-cycle (see figure 2 in \cite{zhang96}) is associated with the pinning
and unpinning of vortices in the superfluid with resulting discrete interaction with the outer crust.
These interactions induce changes in the both the parallel and perpendicular components of the
moments of inertia of the outer crust and excite TFP.
Some of the unpinned superfluid likely rotates with the outer crust up to $\sim1100\Hz$. This
superfluid then repins with the pinned superfuild causing the spin down of the outer crust.
Since the outer crust in an LXMB NS is weakly coupled to the interior of the NS as compared to the crust
of a NS in a radio pulsar, the manifestation of the observed
"glitch" is slow both going up and down in an LMXB NS.

In addition the accretion onto the surface of the outer crust also causes torques which excite TFP. 
The outer crust is also continuously elastically responding to the changes in the internal strain.
These internal adjustments also contribute to fluctuations in the components of the moments of inertia
and the excitation of TFP. 
Simultaneous with all of these effects, GWs are emitted due to TFP.  
Using equation 2.13 from \cite{cutler00} in an entirely different context,
we estimate the exponential timescale
for GW damping of the TFP ($\sim500 s$). Equation \ref{damp} is a recasting of this equation assuming that
($\Delta I \over I$) is of order unity rather that $\sim 10^{-7}$. In the context of the TFP model
the outer crust responds elastically to changes in spin rate completely independent of the shape of the
fluid core of the NS which is spinning slower than the outer crust.
\begin{equation}
\tau_{\theta_p} = 5.4 \times 10^2
               { \left[ {1~{\rm kHz}} \over \nu_{cs} \right] }^4
               { \left[ { 10^{41} {\rm g~cm^2} } \over I_{c\parallel} \right] } s
\label{damp}
\end{equation}
The slower positive and negative variations ($\sim100s$) in the spin of the outer crust
seen in figure 2 of \cite{zhang96} may be due to the varying back reaction of the emission
of GWs due to TFP.

\section{Implications of the Torque Free Precession Model}

In the following sections, we discuss a few of the implications 
of the TFP model including GW emission,
LT precession of material orbiting a NS,
periodicities in the X-ray emission from Type~I bursts,
magnetic dipole radiation emitted at the spin frequency of the crust,
and possible models for GRB based on accretion driven conversion of a LMXB NS to a BH.

\subsection{Generation of Gravitational Waves}

The accretion in the brighter LMXB sources may exceed the Eddington rate.
Examples of mass transfer in LMXB that result in
accretion rates as high as $10^{-5}~\msol~{\rm yr}^{-1}$
have been modeled by \cite{harpaz94} and \cite{iben97}.
Companion stars with masses greater than the NS ($\sim3\msol$) have
stable accretion rates due to Roche lobe overflow on a thermal time
scale that can reach super Eddington levels (\cite{pylyser88}; \cite{rappaport00}). 
The full range of conditions of possible donor stars in accreting
binary systems allows a wide variation in the accretion rate.
As the rate approaches or exceeds the Eddington limit
for a portion of the lifetime of the X-ray source,
the X-ray emitting NS may decrease the
accretion onto the NS surface by radiatively ejecting material from the system.
Even though such complex radiative driven processes may occur,
we suggest that in some cases higher accretion rates are sustained in a
steady state.

%via the dominant and efficient release of energy in the
%form of gravitational waves (GW) due to TFP of the crust of the NS.

%A general discussion of the generation of GWs by a spinning NS is found in
%\cite{thorne87} and references therein.

In particular, we propose that in systems with high accretion rates, 
the NS spins up to near breakup and emits most of the accretion
derived energy in GWs produced by the TFP of the NS crust.  
The validity of the concept of classical TFP in
general relativity (GR) has been demonstrated by \cite{thorne83}.
TFP implies the possibility that the unmeasured emission of GWs may exceed the X-ray
emission. Therefore in the subsequent discussion we consider that some
LMXB NSs may have extreme super Eddington accretion rates only limited
by realistic models for flow of material from donar stars.
GWs emitted due to TFP of sufficient amplitude may cool the accretion
flow allowing a significant fraction of the accreting material
to reach the surface of the NS. Also the rapid rotation of the outer crust
as it approaches contact with the inner edge of the accretion disk at
a matching velocity reduces the conversion of kinetic energy at the surface of the NS.
The X-ray emission from LMXB may be poor indicator of the accretion
rate if significant TFP is present. However, it may be true that
on average the LMXBs with lower X-ray emission (atoll sources) and those
with higher X-ray emission (Z-sources) still have a range of accretion
rates both below and above the Eddington limit. Even sources with current
low accretion rates may undergo large amplitude TFP driven by stored internal
NS energy. We have little evidence concerning the relationship between
the long term average accretion rate and current observed X-ray emission.
The likelihood of an unmeasured emission of GWs from LMXB should caution
any firm interpretations about the total accretion onto the surface of the NS.

The following estimation of the emission of GWs via NS precession is derived
entirely from the work of \cite{zimmermann79} and \cite{zimmermann80}.
The power at the inertial precession frequency $\nu_{cs}+\nu_{cp}$ emitted
by the precessing crust of the NS
in the form of gravitational waves $L_{GW}(\nu_{cs}+\nu_{cp})$ is given by equation \ref{GWp}
derived from \cite{zimmermann79}. We assume that the precession angle $\theta_{cp}$
is small and neglect emission at the second harmonic of the inertial precession
frequency.

\begin{equation}
L_{GW}(\nu_{cs}+\nu_{cp}) = {64{\pi^6} \over 5} {G \over c^5} {\epsilon_c^2} {I_{c\parallel}}^2
         {\left( \nu_{cs}+\nu_{cp} \right)}{^6} {\theta_{cp}^2}
\label{GWp}
\end{equation}

We define $\alpha$ in equation \ref{alpha}
where $I_{c\parallel}$ is the moment of inertia of part of the crust that is precessing,
and $I_{f\parallel}$ is the moment of inertia of the 
fluid interior that spins without precessing (as discussed
in the previous section). For estimation of GW emission, we ignore the frequency dependence
of $\alpha$ implied by equation \ref{Iparallel}.
The parameter $\theta_{cp}$ is the angular amplitude 
of the precession in radians.

\begin{equation}
\alpha = {I_{c\parallel} \over I_{f\parallel}}
\label{alpha}
\end{equation}

Assuming $\alpha$ and $\theta_{cp}~<<~1~~$, we then estimate $L_{GW}(\nu_{cs}+\nu_{cp})$
with selected values of variables in equation \ref{GWestimate}.
The value of $I_{f\parallel}$ is from a NS model with an EOS FPS.
The value of $\nu_{cs}+\nu_{cp}$ is set for the condition of
maximal rotation of the crust near NS breakup. The values of $\alpha$ and $\theta_{cp}$
are not well determined and would be best constrained by future direct observation of the GWs.
Note that the modest values of these two parameters imply that crust TFP is a
very efficient emitter of GWs.
Theoretical NS models for the outer crust suggest that $\alpha\sim10^{-4}$
(\cite{mochizuki97}; \cite{shapiro83}). We assume that the inner crust has melted as
discussed previously.
The values for $\alpha$ and $\theta_{cp}$ are uncertain but are estimated
based on the specific scenario of TFP of the outer crust. Alternative variations of
the scenario would still alllow TFP with different values for $\alpha$ and $\theta_{cp}$.
For TFP in an LMXB NS we select a value of $\theta_{cp}\sim0.02-0.1$ based on the observations of
the amplitude of the lower kHz QPO (RMS of the mean X-ray flux; \cite{klis96}).
The ease of detection of the
kHz QPO in LMXB suggests significant amplitudes for $\alpha$ and $\theta_{cp}$.
The lower values of $\theta_{cp}$ are likely associated with higher values of $\nu_u$
because of the increasing GW emission at higher spin rates. 

The modulation of the X-ray emission at the precession frequency $\nu_{cp}$
($=\nu_{\ell}$) could be caused either by geometrical effects similar to those for X-ray pulsations
from a spinning NS or because the
velocity and acceleration vary at the surface of the NS modulated by TFP
with an amplitude proportional to $\theta_{cp}$.
These variations on the surface of the NS would
modulate the accretion energy and thereby be tracked by
modulation in the X-ray emission on timescales of order $\sim1~{\rm ms}$.
Justification for the rapid tracking of these modulations is implied by the radiation
dominated conditions on the surface on the NS (\cite{arons92}; \cite{klein96}; \cite{jernigan00}).

The energy emitted in GW at the inertial crust precession frequency seen by a distant observer
is given by \ref{GWestimate}.

\begin{equation}
L_{GW}(\nu_{cs}+\nu_{cp}) = 2.3\times 10^{42}
           {\left[ \epsilon_c \over 0.7 \right]}^2
	   {\left[ I_{f\parallel} \over 1.2\times 10^{45}~{\rm g~cm^2} \right]}^2
           {\left[ {\nu_{cs}+\nu_{cp}} \over 1700\Hz \right]}^6
           {\left[ \alpha \over 10^{-4} \right]}^2
	   {\left[ \theta_{cp} \over 0.02~{\rm radians} \right]}^2
	   {\rm ergs~s}^{-1}
\label{GWestimate}
\end{equation}

Equation \ref{GWspin}
(derived from the first term of equation 14 in \cite{zimmermann80})
is an estimate of the gravitational power emitted at the second harmonic of the
spin frequency of the crust due to fluctuations in the components of the moments of inertia
of the NS. Here we estimate the fluctuations of $I_{c\perp}$
as $\sim I_{c\perp}/Q$ where Q is a typical quality factor of the upper kHz QPO.
Previously we simplified the TFP model by assuming that both components of $I_{c\perp}$
were equal. Here we must assume that the two orthogonal components of $I_{c\perp}$ can
fluctuate independently.

\begin{equation}
L_{GW}(2 \nu_{cs}) = 1.6\times 10^{42}
	   {\left[ I_{f\parallel} \over 1.2\times 10^{45}~{\rm g~cm^2} \right]}^2
           {\left[ \nu_{cs} \over 1000\Hz \right]}^6
	   {\left[ Q \over 100 \right]}^{-2}
           {\left[ \alpha \over 10^{-4} \right]}^2
	   {\rm ergs~s}^{-1}
\label{GWspin}
\end{equation}

For Sco~X-1 at a distance of 2.8 kpc (\cite{bradshaw99}) and with the selected values
for $\theta_p$ and $\alpha$, the amplitude of the GWs
emitted at frequency $\nu_{cs}+\nu_{cp}$ is given in equation \ref{GWh} and
numerically estimated in \ref{GWnumerical}.

\begin{equation}
h_{GW}(\nu_{cs}+\nu_{cp}) = {4 \pi^2}{G \over c^4}~{{I_{c\parallel} (\nu_{cs}+\nu_{cp})^2} \over d}~{\alpha \theta_{cp}}
\label{GWh}
\end{equation}

\begin{equation}
h_{GW}(\nu_{cs}+\nu_{cp}) = 7.5\times~10^{-24}
    {\left[ d \over {2.8~{\rm kpc}} \right]}^{-1} 
    {\left[ \nu_{cs}+\nu_{cp} \over 1700\Hz \right]}^2 
    \left[ \alpha \over 10^{-4} \right] 
    \left[ \theta_{cp} \over 0.02 \right] 
\label{GWnumerical}
\end{equation}

Such an amplitude would indicate that the
advanced LIGO system ( $h_{3/yr}~\sim~10^{-26}$ at $1700\Hz$, \cite{brady98})
could detect this GW emission within a single day of integration.
Such an attempt would require simultaneous tracking of the kHz QPO in X-rays
so that power density spectra of the GW emission could be coherently
coadded with appropriate frequency shifts. Equation \ref{heff} is an estimate
of the amplitude of the GW emission at the sensitivity threshold for an observation with the
advanced LIGO of duration $T_{obs}$ where Q is the quality factor of the lower kHz QPO.

\begin{equation}
h_{eff}(\nu_{cs}+\nu_{cp}) = 3.4\times~10^{-24}
    \left[ h_{3/yr} \over 10^{-26} \right] 
    {\left[ Q \over 100 \right]}^{-{1 \over 4}} 
    {\left[ \nu_{cs}+\nu_{cp} \over 1000\Hz \right]}^{1 \over 4} 
    {\left[ T_{obs} \over 1~{\rm day} \right]}^{-{1 \over 4}} 
\label{heff}
\end{equation}

Equation \ref{heff} shows the weak dependence on $T_{obs}$ 
expected for the addition of many power density spectra for periodic GW emission
with a finite Q. X-ray detectors with apertures much larger than the PCA onboard
RXTE (\cite{bradt93}; \cite{swank97}) could track individual kHz oscillations. Such X-ray data could
be combined with LIGO data to achieve much better sensitivity. 
Even without such detailed X-ray data,
the prospects for LIGO detection of GWs from LMXB sources and Sco~X-1 in particular are possible
though uncertain.

\subsection{Gravitational Waves at Other Frequencies}

The dominant frequencies of expected
GWs ($\nu_{cs}+\nu_{cp}=\nu_u+\nu_{\ell}$ and $2\nu_{cs}=2\nu_u$) are generated
from precession and rotation of the outer crust respectively.
Measurements of GWs at these frequencies would determine the magnitude of $\alpha$ and $\theta_{cp}$
and confirm the estimates for these parameters based on the X-ray
emission and thereby constrain the details of the precessional motion.

This motion will likely
excite distinct modes corresponding to spherical harmonics of the NS (\cite{ostriker73}).
Some of these modes can induce a triaxial form of either the crust or fluid interior of the NS.
This would cause the emission of GWs at frequency $2\nu_u$ from the outer crust (see equation \ref{GWspin})
and at frequency $2\nu_{fs}$ from the fluid interior
(nearly constant value for a particular LMXB NS in the range $300-800\Hz$).
These motions could also drive the g-mode ocean waves on the surface of the crust (\cite{bildsten96})
or modify r-mode oscillations present in the fluid interior (\cite{owen98}; \cite{bildsten00}).
In both of these cases the influence of near breakup rotation and the precession of the NS
may significantly alter predictions from previous work.
Since the precession and spin frequencies vary over a wide range in a single LMXB,
there is the possibility that some of these modes
will resonantly match the precession or spin frequencies or their harmonics.
Future observations of a full GW spectrum with many frequencies present, including the effects of
temporal evolution of these modes in response to changes in the precession frequency,
will constrain many details of NS structure beyond just the EOS, including fluid properties,
vortex pinning, viscosity, internal flows and crust and core structure. 

\subsection{Lense-Thirring Precession}

Following the work of \cite{stella98}, we revise the possibility of
the presence of Lense-Thirring (LT) and classical precession of material orbiting near the NS
in the context of the TFP model.
The total precession frequency can be evaluated
as a sum of two terms $\nu_{LT}$, the LT precession, and $\nu_{cl}$, the classical precession.
These two terms have contributions from the rotating fluid core and the
decoupled crust which are each spinning at different rates. The total LT
effect of the rotating crust is small since the moment of inertia
$I_{c\parallel}$ ($=\alpha I_{f\parallel}$)
is small ($\alpha << 1$).
We estimate $I_{f\parallel}=1.3\times10^{45}$~g~cm$^2$ at rotation rate $\nu_{fs}\sim740\Hz$
using the NS model based on EOS FPS from \cite{cook94b}.
The LT term for the NS fluid core is given by equation \ref{LT}
and numerically estimated by equation \ref{LTestimate}.

\begin{equation}
\nu_{LT}~ = ~ {{8 \pi^2 I_{f\parallel} {\nu_k}^2 \nu_{fs}} \over c^2M}
\label{LT}
\end{equation}

\begin{equation}
\nu_{LT}~ = ~45
	\left[ I_{f\parallel} \over 1.3 \times 10^{45}~{\rm g~cm}^2 \right]
	{\left[ M \over 1.4~\msol \right]}^{-1}
	\left[ \nu_k \over 1200\Hz \right]^2
        \left[ \nu_{fs} \over 740\Hz \right] \Hz
\label{LTestimate}
\end{equation}

where $M$ is the mass of the NS and $\nu_k$ is the Keplerian orbital
frequency of the material that is precessing.

The classical precession effect is given by equation \ref{cl} and numerically estimated
by equation \ref{clestimate}.

\begin{equation}
\nu_{cl} ~=~ {3 \over {8 \pi^2}} { { G \cos \beta } \over { r^5 \nu_k } } {\left( I_{f\perp} - I_{f\parallel} \right)}
         ~=~ -{3 \over {16 \pi^2}} { { G \cos \beta } \over { r^5 \nu_k } } { {e_f^2} \over {1 - e_f^2} } I_{f\parallel}\quad
\label{cl}
\end{equation}

\begin{equation}
\nu_{cl} ~=~ -2\times10^{-2}~ {\left[ I_{f\parallel}  \over 1.3\times 10^{45}~{\rm g~cm^2} \right]} 
                    {\left[ M \over 1.4~\msol \right]}^{-{5 \over 3}} 
                    {\left[ \nu_k \over 1000\Hz \right]}^{ 7 \over 3}
                    \cos \beta \Hz
\label{clestimate}
\end{equation}

In the above equation for $\nu_{cl}$ we set the value of $e_f=0.4$ from model FPS
for a spin rate $\nu_{fs}\sim740\Hz$ (see table 15 in \cite{cook94b}).
The ellipticity $\epsilon_f$ and eccentricity $e_f$ are defined for the fluid core 
in analogy to equation \ref{epsilon_c} for the crust, 
where $I_{f\parallel}$ and $I_{f\perp}$ are the moments of 
inertia of the fluid core of the NS.  

\begin{equation} \epsilon_f~=~{e_f^2 \over {2~-~e_f^2}}~=~
{{\left[ I_{f\parallel} - I_{f\perp} \right]} \over {I_{f\perp}}}
\label{ef}
\end{equation}
Equation \ref{clestimate} shows that classical precession negligible
compared to LT precession. 

These estimates are consistent with the basic LT conjecture of \cite{stella98}
except that their assumption that $\nu_k~=~\nu_u$ must be based on a indirect correlation. We suggest
that both $\nu_k$  and $\nu_u(=\nu_{cs})$ are independently proportional to the accretion rate, $\dot{M}$.
Another possibility is that the precession of the crust at 
frequency $\nu_{cp}$ ($=\nu_{\ell}$) induces a tilt in the
accretion disk at the location where material orbits the NS at the TFP frequency
(not to be confused with the precession of orbiting material).
This in turn leads to a modulation of the accretion rate by material in orbit at the frequency
$\nu_k~=~\nu_{cp}$. This would provide a reason for identifying $\nu_k$ with $\nu_{\ell}$
yielding a similar result to that in \cite{stella98} with a somewhat different interpretation.

The prior difficulty in \cite{stella98} of a factor of $\sim4$ 
discrepancy for the $I \over M$ for a NS may be
removed if NS spin frequency $\nu_{fs}$ is as high as $\sim800\Hz$.
In the context of the TFP model, the spin of the fluid core $\nu_{fs}$ is not related
to the difference frequency of the kHz QPO. For example in the source Sco X-1, $\nu_{fs}$
may be as high as the minimum observed value of $\nu_u$ ($\sim845\Hz$;\cite{klis00}).
Setting the value of $\nu_{fs}\sim740\Hz$ in the equation \ref{LTestimate} yields
a $\nu_{LT}\sim45\Hz$ in agreement with observations.
We are not proposing a specific detailed version of the effects of LT precession but only
caution that both $\nu_{fs}$ and $\nu_k$ are uncertain. 

%Equation \ref{LT} should really be modified to add $I_{c\parallel}$ to $I_{f\parallel}$ since
%there is a contribution to the LT frequency from the crust. This component of the LT frequency
%is dependent on the spin rate of the crust which is identified as $\nu_u$.
%The change would slightly modify the simple proportionality $\nu_{LT}\sim{\nu_k}^2$
%assuming that ${\nu_k}={\nu_{cs}}={\nu_u}$
%(or the alternative ${\nu_k}={\nu_{cp}}={\nu_{\ell}}$).
%The same necessary modification to equation \ref{cl}
%does not change the fact that the classical precession
%is small compared to LT precession.

\subsection{Periodicity Observed in Type~I Bursts}

Observations indicate a near coincidence of $\Delta\nu$ ($=\nu_u-\nu_{\ell}$) with $\nu_{burst}$
or $\nu_{burst}/2$. The conclusion that $\nu_{burst}$ is near the actual spin rate of the
surface of the NS is inconsistent with the simple version of the TFP model since
$\nu_{cs}$, the spin frequency of the NS surface, is significantly larger than $\Delta\nu$.
The TFP model is consistent with an rapidly rotating outer crust
(up to $1100\Hz$ with eccentricity $\sim0.93$)
which is flatten in form compared to the nearly spherical rotating fluid core
($300-800\Hz$ with eccentricity $\sim0.2$). 
The NS in Sco~X-1 may possibly be rotating as fast as $\sim800\Hz$.
In such a configuration the accretion rate onto the equatorial region may be
significantly higher than onto the polar regions.
This possibility is consistent with the lack of any detection of oscillations at the frequency
$\nu_{fs}$ in the X-ray emission due to steady accretion.
We suggest the following possibility
that the outer crust is rotating with the fluid core in the polar regions ($300-800\Hz$)  
and is rotating independently and more rapidly in the equatorial region (up to $1100\Hz$).
Further in the mid-latitude regions there is the possibility that the outer crust
rotates with an internal superfluid layer at a rate $\sim5\Hz$ faster than the fluid core.
Unstable thermonuclear burning would occur for each latitude region
depending on the level of the local accretion rate.
The progression of observed frequencies in a Type~I burst would depend on the particular
pattern of the burning front. The scenario is consistent with the frequencies observed in
Type~I bursts (see \cite{strohmayer98}).
The fractional variation in the spin rate of the fluid core $\nu_{fs}$ as the crust
spins up and down is $\sim10^{-4}$ assuming full transfer of the angular momentum between
the outer crust and the fluid core of the NS. This result is consistent with the
repeatability of the asymptotic NCO frequency observed in several independent Type~I bursts
from 4U1728-34 and 4U1636-53
($1.5\times10^{-4}$ and $2.3\times10^{-4}$ respectively; \cite{strohmayer98}).
In SAX~J1808.4-3658, we expect higher Q NCO at the spin frequency ($401\Hz$) which is
the same for both the crust and core of the NS.

Recently \cite{wijnands00} discovered nearly
coherent oscillations (NCO) in Type~I bursts from X1658-298. These NCO occurred at
frequencies $\sim567\Hz$ during the main part of the burst
and at frequency $\sim572\Hz$ during the tail of the burst.
This pattern of frequencies is consistent with a burning front which migrates from
a polar region to a mid-latitude regions. The discrete jump in frequency matches
that expected for the superfluid layer that spins differentially from the NS core
as discussed previously. 
The burning front may continue into the equatorial region which might produce
a feature in the power density spectrum similar to the upper kHz QPO.
The higher frequency of such a peak and its finite Q may decrease the likelihood of
detection. Alternately, higher acccretion rates in the equatorial region may suppress
unstable thermonuclear burning.

\subsection{Magnetic Dipole Radiation}

The maximum spin rate of the surface of a NS is the maximum observed value of the upper kHz QPO.
Such high spin rates suggest that LMXB NSs are
candidates for the direct detection of magnetic dipole radiation since
the power emitted is strongly dependent on spin rate. The following equation
is an estimate of the power emitted from Sco~X-1 where $B_0$ is the surface magnetic field, $r_e$ is
the equatorial radius of the NS, and ${\theta}_d$ is the viewing angle of a distant observer
relative to the spin axis.
The numerical estimate of the power emitted given by equation \ref{dipole}
assumes a viewing angle normal to the rotation axis 
($\theta_d = {\pi \over 2}$)

\begin{equation}
P_{dipole}~ = { { {r_e}^6 {B_0}^2 \sin^2 {\theta}_d } \over {6 c^3} }
        = 6.2\times 10^{38}
         {\left[ r_e \over 2 \times 10^{6}~{\rm cm} \right]}^6
         {\left[ B_0 \over 10^{9}~{\rm Gauss} \right]}^2
         {\left[ \nu_{cs} \over 1000\Hz \right]}^4 {\rm ~ergs~s}^{-1}
\label{dipole}
\end{equation} 

This radiation, if present, is energetically of the order of the X-ray portion of the accretion energy 
but likely less than the energy emitted in GWs.
The plasma near the NS might convert much of this energy into local
plasma modes; however the wavelength of the radiation ($\sim 10^7$ cm) exceeds
the size of the NS by a large factor, therefore some of this power may leave the system.

The plasma dynamics near the NS and interaction with the interstellar medium along
the line of sight will likely decrease the chances of detection of this radiation.
However since Sco~X-1 is nearby on a galactic scale (2.8 kpc), perhaps detection
is possible by spacecraft such as the Voyagers (\cite{gurnett96}), now located
in the outer solar system where the plasma frequency is sufficiently low.
The plasma frequency $\nu_{pl}$ is related to the local electron density by:

\begin{equation}
\nu_{pl} ~=~ {\left[ { e^2 N_e } \over {\pi m_e} \right] }^{1 \over 2}
        = 8.97\times 10^3 {N_e}^{1 \over 2}\Hz
\end{equation} 

If at any point along the line of sight the electron density rises above
$\sim10^{-2}$ cm$^{-3}$ then such plasma will damp out the direct dipole radiation
at frequency $\nu_u$ ($\sim1000\Hz$).
The solar wind drops below this density in the outer regions of the solar system. Also
the Voyagers are within about a decade of reaching the heliopause (see figure 7 of \cite{gurnett96}).
There is also the possibility of electromagnetic radiation at somewhat higher frequency
due to TFP ($\nu_{cs}+\nu_{cp}~\sim~1700\Hz$). 

%The electric field detectors onboard Voyager are able to detect fields as small
%as a few mV m$^{-1}$ which is sensitive enough to 
%detect the direct dipole radiation from Sco~X-1 if it is present
%($\sim??~{\rm mV~m$^{-1}$ assuming a distance of 2.8 kpc and power estimated by equation \ref{dipole}).

\subsection{Progenitors of Gamma Ray Burst Sources}

Neutron stars in LMXB spinning near breakup speed with high accretion rates
would likely emit most of their energy in the form of GWs while the X-ray emission is maintained
at levels influenced by the Eddington limit.
The LMXB sources with low mass donor stars may be the progenitors of
millisecond radio pulsars (\cite{radhak82}; \cite{alpar82}).
If the donor companion star is massive enough (few $\msol$) to provide
$\sim0.5~\msol$ of material, then the NS might approach the maximum stable mass while rotating
near the breakup rate.
The GW cooling of the accretion flow near the surface of a NS may allow a large
fraction of the material that leaves the companion star to reach the surface of the NS.
The X-ray emission is still restricted by the Eddington limit but at much higher
mass accretion rates onto the NS surface.
The fluid core would collapse into 
a maximally rotating Kerr black hole ($\sim2~\msol$) and
perhaps release energy of a fraction of $10^{54}$ ergs
in the form of jets of material, GWs, neutrinos, and gamma-rays.
An LMXB NS could be a progenitor of a GRB 
which leaves behind a maximally spinning black hole (\cite{cui98}).
Subsequent continued accretion from the donar companion star would increase the BH mass.
To explain the observed properties of the brighter GRBs would likely require conversion
of not just the kinetic energy of the NS material as it forms the black
hole (BH) but conversion of some of the rest mass of the NS into gamma-rays. 

\cite{ruffini99} has suggested that temperatures in the dyadosphere
of the BH at the time of formation may rise enough
to cause copious $e^{\pm}$ pair creation and subsequent annihilation with conversion to gamma-rays.
As much as $50\%$ of the mass-energy of the NS
might be released at the time of formation of a maximally spinning Kerr BH (\cite{ruffini99}).
After the BH forms, the crust of the NS ($\sim10^{-4}~\msol$) may still be present and
spinning at frequency $\nu_{cs}\sim1000\Hz$.
The remnant crust will no longer have pressure support
and will likely transform to a triaxial form which will quickly form two blobs of
rotating hot plasma around the BH at a radius of the former NS.
This material ($\sim10^{-4}~\msol$) will fall onto the BH releasing
its kinetic energy in the form of GWs, neutrinos, electromagnetic waves and jets of material.
We crudely estimate the time for this remnant NS crust to accrete onto the BH due to emission of GWs
as the ratio of the sum of kinetic ($E_k$) and potential ($E_p$) energies to the luminosity emitted
in GWs ($L_{GW}$). This ratio is $\left( { E_{ke}+E_{pe} } \over {L_{GW}} \right) \sim1~$s from
equations \ref{GWcrust}, \ref{GWke} and \ref{GWpe}. This estimate is likely too fast since
$L_{GW}$ is probably an upper limit and release of other forms of energy will likely retard the flow
towards the BH event horizon. 
This model for a GRB might have a unique signature of nuclear composition since the accreting
material is from the breakup of the 
remnant crust of a NS (\cite{lattimer77}).
The spectrum of the gamma-ray emission could in principal reveal the details of the nuclear
composition (\cite{hailey99}).

\begin{equation}
 L_{GW}~ = ~{512 \pi^6 G \over 5 c^4} I_{c\parallel} {\nu_{cs}}^6
        = 5.2\times 10^{49}
         \left[ I_{c\parallel} \over 2.4 \times 10^{14}~{\rm g~cm}^2 \right]
         {\left[ \nu_{cs} \over 1000\Hz \right]}^6
          {\rm ergs~s}^{-1}
\label{GWcrust}
\end{equation} 

\begin{equation}
E_{k}~ = ~{2{\pi}^2}I_{c\parallel}{\nu_{cs}}^2
        = 4.7 \times 10^{49}
          \left[ I_{c\parallel} \over 2.4 \times 10^{41}~{\rm g~cm^2} \right]
          {\left[ \nu_{cs} \over 1000\Hz \right]}^2
          {\rm ergs}
\label{GWke}
\end{equation} 

\begin{equation}
E_{p}~ = ~ {{G M M_c} \over r_e}
        = 3.3 \times 10^{49}
          \left[ M \over 1.4~\msol \right]
          \left[ M_c \over 10^{-4}~\msol \right]
          {\left[ r_e \over 10~{\rm km} \right]}^{-1}
          {\rm ergs}
\label{GWpe}
\end{equation} 

In rough numbers we can estimate the rate of GRBs that would be produced by accreting
$\sim0.5\msol$ of matter onto the NS in an LMXB and thereby initiating its conversion
to a BH. Assuming a long term average accretion rate of $10^{-7}~\msol~{\rm yr}^{-1}$ 
yields a rough lifetime for each LMXB, $t_{LMXB} \sim 10^7$ yr.
If the universe contains $10^{11}$ galaxies equivalent to the Milky Way,
and each galaxy contains a few LMXB ($N_{LMXB}$) at any time which are accreting at high
rates from donor companion stars of a few solar masses, and the radiation is beamed
by a fraction $f_b$ 
then the total rate of observed GRB ($R_{GRB}$) is estimated in equation \ref{GRB}.

\begin{equation}
R_{GRB}~ \sim 1
	      \left[ f_b \over 10^{-2} \right]
	      \left[ N_{gal} \over 10^{11} \right]
              \left[ N_{LMXB} \over 3 \right]
              {\left[ t_{LMXB} \over 10^{7} {\rm yr} \right]}^{-1} {\rm  day}^{-1}
\label{GRB}
\end{equation}

\section{Conclusions}

The basic theory of TFP of a NS as the signature of
the approach to NS breakup is a viable explanation of the kHz QPO
observed in X-rays emitted by LMXB sources.
A simple version of the theory relates the intrinsic properties
of NS structure with the observed kHz frequences.
The range of kHz frequencies observed is also
explained by this simple dynamical model.
Furthermore, the TFP theory provides a simple
explanation for the high Q of the millisecond pulsar SAX~J1808.4-3658 including the
reason why it does not exhibit kHz QPO.
Since the theory relates the ratio of the observed kHz frequencies to the
ratios of the moments of inertia of the NS, it is not
surprising that the EOS of NS matter is so tightly constrained by this interpretation.

The primary issue not fully addressed by the TFP model
is the complex interaction of the crust and core of the NS.
On a short timescale ($\sim$seconds) the fluctuations in 
the kHz QPO frequencies are explained by
likely fluctuations in the moments of inertia of the NS crust.
On a longer timescale ($\sim$hours) internal torques  drive the system into a limit-cycle
with a range of crust spin frequencies. Angular momentum may be
conserved since the three dominant and independent reservoirs of angular momentum
are the rotation of the fluid core, the rotation of the crust, and precession of the crust.
These components can exchange angular momentum and vary their moments of inertia.
The simple TFP model constrains the locus of the ratios of the kHz QPO frequencies
but does not explain all the details of the dynamics.
We have also outlined a plausible mechanism for a limit-cycle of exchange of angular
momentum between the outer crust and a superfluid layer internal to the NS
that is analogous to giant glitches in radio pulsars.
The TFP theory predicts strong kHz GW emission and magnetic
dipole radiation as very low frequency radio waves ($\sim1000\Hz$).

A LIGO detection of GWs from an LMXB NS source due to TFP would be the 
definitive confirmation of the TFP theory.
The high sustained accretion rates that such
GW emission would require suggests that some LMXB NSs will collapse into BHs.
This possibility suggests that NSs in LMXB are progenitors of some GRBs and
leave behind maximally rotating Kerr BHs as the final successors to these GRBs.

\section{Acknowledgments}

The author acknowledges Lynn Cominsky for helpful conversations
and a careful reading of the manuscript.
The author also thanks Michiel van der Klis for providing the previously
published Sco~X-1 data on kHz QPO in numerical form.
Walter Lewin, Saul Rappaport, Wei Cui and Ben Owen are acknowledged
for their generous efforts reading and commenting on the manuscript.
NASA provided the primary support for the author who is 
a Co-I on the MIT portion of the RXTE project (PI: Hale Bradt)
under a subcontract to the University of California, Berkeley (NAG 5-30612).
The author thanks Hale Bradt for his sustained support of this subcontract.
The author also thanks Bill Mayer and George Clark, the project scientist 
and principal investigator for SAS-3 respectively, and the APL engineers
that build the spacecraft for SAS-3 for providing the opportunity to
study torque free precession of coupled rigid bodies. In particular,
the nutation damper crisis that occurred just after the launch of SAS-3
seeded the concept for this paper. 
Additional support was also provided by Eureka Scientific (President: John Vallerga).
The work was carried out at the Space Sciences Laboratory of the University of California, Berkeley,
and at the Little H-Bar Ranch located in Petaluma, California.

\clearpage

\figcaption{ (a) the difference frequency $\nu_u-\nu_{\ell}$
versus the Upper kHz QPO frequency $\nu_u$ for Sco X-1;
The single point for GX~5-1 is the observation with the lowest value of $\nu_u$.
The solid curve is the best fit TFP model. The dashed curve is the 
phenomenological function derived by \cite{psaltis98};
(b) an expanded version of (a) showing the Sco X-1 data with error bars and the
two model fits;
(c) the ratio $\nu_{\ell}/\nu_u$ versus $\nu_u$.
The TFP model is the solid curve;
The dashed curve is the locus for a MacLaurin oblate spheroid.
The locus is shown a model of a NS based EOS FPS from \cite{cook94b}.
A vertical dot-dash line (label J1808) marks the spin frequency ($401\Hz$) of SAX~J1808.4-3658.
The three critical values of eccentricity $e$ are marked by horizontal dotted lines;
(d) an expanded version of (c) showing the Sco X-1 data in detail.
}

\begin{figure}
\plotone{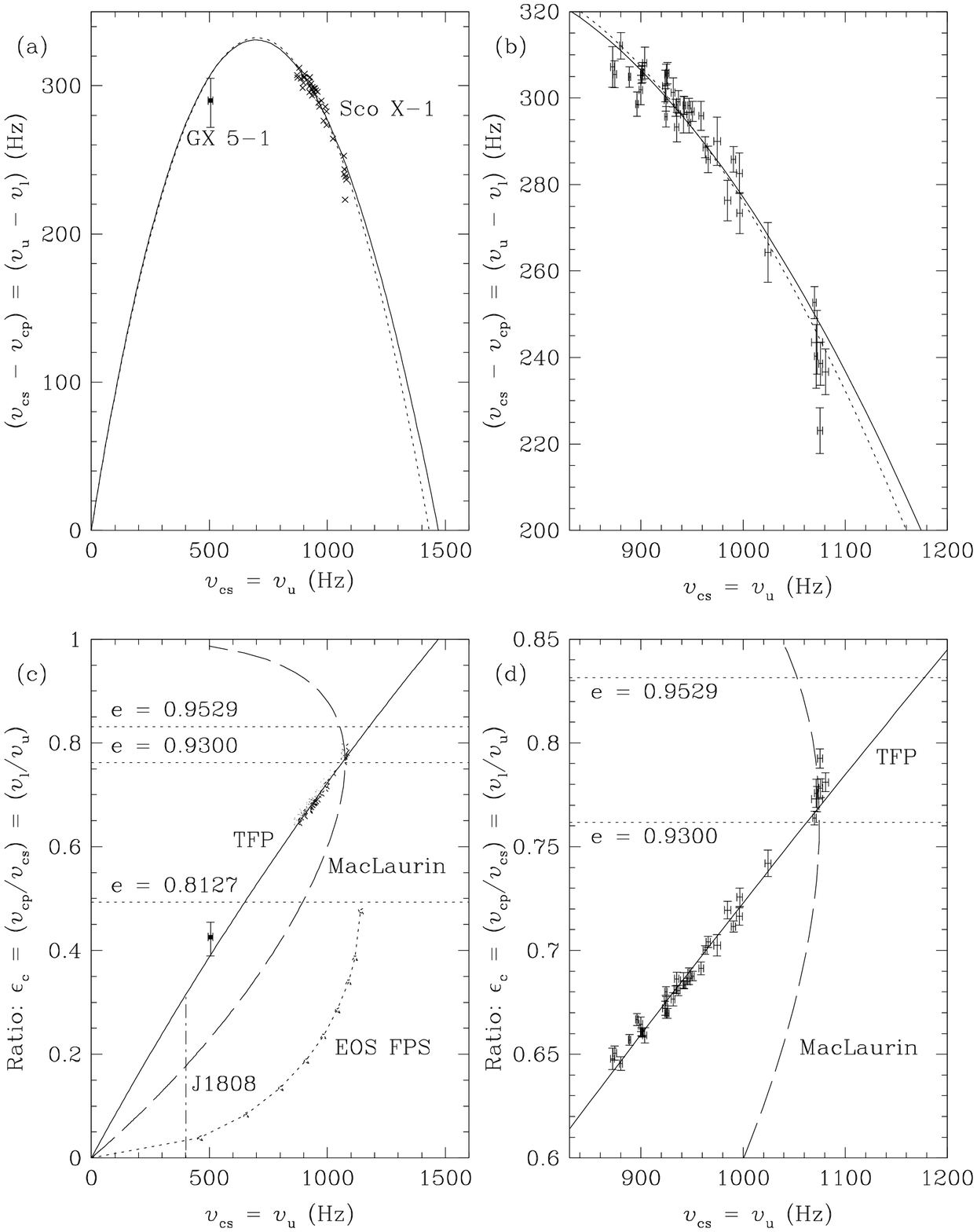}
\label{figscox1}
\end{figure}

\end{document}